\documentclass[namedreferences,hyperref,optionalrh]{spr-sola}
\usepackage{graphicx}        % For eps figures, newer & more powerfull
\usepackage{amssymb}        % useful mathematical symbols
\usepackage{color}           % For color text: \color command
\usepackage{amsmath}
\usepackage{tikz}
\renewcommand{\vec}[1]{{\mathbfit #1}}

\newcommand{\dl}{~{\mathrm d} l}

\newcommand{\XX}{\vec X}
\newcommand{\BB}{\vec B}
\newcommand{\CC}{\vec C}
\newcommand{\bb}{\vec b}
\newcommand{\EE}{\vec E}
\newcommand{\FF}{\vec F}
\newcommand{\jj}{\vec j}
\newcommand{\RR}{\vec R}
\newcommand{\xx}{\vec x}
\newcommand{\uu}{\vec u}
\newcommand{\ee}{\vec e}
\newcommand{\bsigma}{\boldsymbol{\Sigma}}
\newcommand{\bOmega}{\boldsymbol{\Omega}}

\chardef\us=`\_

\begin{document}

\begin{frontmatter}
\title{On field line slippage rates in the solar corona}

\author[addressref={aff1},email={david.mactaggart@glasgow.ac.uk}]{\inits{D.}\fnm{David}~\snm{MacTaggart}\orcid{0000-0003-2297-9312}}

%\author{\inits{}\fnm{}~\lnm{}\orcid{}}
%   NOTE:  Just one corresponding author [corref]
\address[id=aff1]{School of Mathematics and Statistics, University of Glasgow, Glasgow, G12 8QQ, UK}

\runningauthor{MacTaggart}
\runningtitle{On field line slippage in the solar corona}

\begin{abstract}
Magnetic reconnection is one of the fundamental dynamical processes in the solar corona. The method of studying reconnection in active region-scale magnetic fields generally depends on \emph{non-local} methods (i.e. requiring information across the magnetic field under study) of magnetic topology, such as separatrix skeletons and quasi-separatrix layers. The theory of General Magnetic Reconnection is also non-local, in that its measure of the reconnection rate depends on determining the maxima of integrals along field lines. In this work, we complement the above approaches by introducing a \emph{local} {description} of magnetic reconnection, that is one in which information about reconnection at a particular location depends only on quantities at that location. {This description} connects the concept of the \emph{field line slippage rate}, relative to ideal motion, to the underlying local geometry of the magnetic field characterized in terms of the Lorentz force and field-aligned current density. It is argued that the dominant non-ideal term for the solar corona, discussed in relation to this new {description}, is mathematically equivalent to the anomalous resistivity employed by many magnetohrdrodynamic simulations. However, the general application of {this new approach} is adaptable to the inclusion of other non-ideal terms, which may arise from turbulence modelling or the inclusion of a generalized Ohm's law. The {approach} is illustrated with two examples of coronal magnetic fields related to flux ropes: an analytical model and a nonlinear force-free extrapolation. In terms of the latter, the slippage rate corresponds to the reconnection which would happen if the given (static) force-free equilibrium were the instantaneous form of the magnetic field governed by an Ohm's law with non-ideal terms.

\end{abstract}
\end{frontmatter}
%-------------------------------------------------

\section{Introduction}
     \label{S-Introduction}

Magnetic reconnection is a fundamental property of plasmas that describes how the topology of magnetic field connectivity changes \citep{pontin2022}. In applications to the solar corona, the key locations of dynamically-important magnetic reconnection (hereafter, reconnection) have been identified through the mapping of topological regions of a given magnetic field. One approach to achieving this is to build \emph{separatrix skeletons} \citep{longcope2005,haynes2010}. Such constructions rely on the existence of null points in order to compartmentalize a given magnetic field into topologically distinct regions. The resulting dome-like separatrix surfaces, connected by null points and separator field lines, represent discontinuous jumps in field line connectivity and are, therefore, the primary locations of where reconnection could take place that changes the global magnetic topology of the given magnetic field. 

It may be the case, however, that a given magnetic field does not contain null points. In this situation, discrete separatrix layers cannot be defined and need to be replaced by a generalization of this concept, resulting in \emph{quasi-separatrix layers} \citep[QSLs; ][]{demoulin1996, titov2002}. QSLs are defined as regions of the magnetic field for which the field line mapping possesses very strong, although continuous, gradients. In this sense, separatrices can be thought of as the singular limit of QSLs. 

Separatrices and QSLs have been successfully combined with the theory of \emph{General Magnetic Reconnection} \citep[GMR; ][]{schindler1988,hesse1988}, which was developed to describe reconnection in magnetic fields without null points. For example,  GMR has been studied in connection with separator reconnection \citep{parnell2024} and close agreements between the predictions of where reconnection occurs according to GMR and QSL analysis have been identified \citep{titov2009}.  One common feature of separators, QSLs and GMR is that they are all \emph{non-local} forms of analysis. Separators and QSLs require the integration of field lines across the entire magnetic field under study, and in GMR the reconnection rate is determined by the integration along field lines of the component of the electric field parallel to the magnetic field. 

In this work, we complement the above theories by presenting a \emph{local} {description} of reconnection, that is one in which the key properties of reconnection can be determined by only considering local information at the place where reconnection is measured. We bring together some existing theoretical results to produce a {description} of reconnection that provides information about the location, strength and direction of reconnection, together with how the reconnection is related to the local geometric properties of the magnetic field through the Lorentz force and field-aligned currents. This {description} provides a powerful tool for analyzing the dynamic behaviour of magnetic fields in the solar corona.  

The layout of the article is as follows. First, we outline {this new description of reconnection} and describe its properties. Secondly, we apply it to two illustrative examples: an analytical model of flux rope formation and a nonlinear force-free (NLFF) extrapolation before the onset of a coronal mass ejection (CME). The paper concludes with a summary and discussion.

\section{Reconnection and field line slippage}
Before discussing reconnection in detail, we must define precisely what we mean by this term. In this work, we adopt a general definition of reconnection \citep[e.g.][]{axford1984} which is also aligns with GMR: if a parcel of plasma initially on one field line becomes connected to another field line, we say that reconnection has occurred. In this sense, the terms \emph{reconnection} and \emph{slippage} are considered to be interchangeable.

In order to provide some context, we now present a very brief overview of the key result of GMR. Given a general Ohm's law of the form 
\begin{eqnarray}\label{Ohm_general}
    \EE + \uu\times\BB = \RR,
\end{eqnarray}
where $\EE$ is the electric field, $\uu$ the plasma velocity field, $\BB$ the magnetic field and $\RR$ represents the non-ideal terms, an important quantity in GMR is the \emph{quasi-potential}
\begin{equation}\label{GMR_rate}
    \Xi = -\int_{\mathcal C} \RR\cdot\bb\dl =-\int_{\mathcal C} \EE\cdot\bb\dl,
\end{equation}
where $\bb=\BB/|\BB|$ is the unit vector along a field line and $\mathcal{C}$ is the curve traced out by a field line. Consider an isolated diffusion region $D\subset\mathbb{R}^3$ defined by
\begin{equation}
    \RR\,\left\{\begin{array}{cc}
        \ne \boldsymbol{0} & {\rm if}\, \xx\in D ,  \\
         = \boldsymbol{0} & {\rm if} \,\xx\notin D.
    \end{array}\right.
\end{equation}
It can be shown \citep[e.g.][]{hesse2005} that in such a region, the reconnection rate $R_D$ is given by

\begin{equation}
    R_D = -\max_D\Xi = \max_D\int_{\mathcal C} \RR\cdot\bb\dl
\end{equation}
This measure of the reconnection rate is non-local because it requires information along an entire field line and not at a single point. It also depends on the maximum of a set of values of $\Xi$. For situations where there are multiple diffusion regions, several local maxima of $\Xi$ would need to be considered \citep{wyper2015}.

Although GMR has been applied successfully to many applications in solar physics, \cite{eyink2015} argues that the rate in equation (\ref{GMR_rate}) should be changed when describing turbulent reconnection. Very briefly, the problem that arises is that in the turbulent inertial range, the magnitude of $\RR$ may be negligible but field line slippage may still occur rapidly throughout the plasma. Although the magnitude of $\RR$ may be negligible, its curl need not be. Thus \cite{eyink2015} proposes an alternative measure of the reconnection rate, the so-called \emph{slip velocity source},
\begin{equation}\label{slip_source_velocity}
\bsigma = -\frac{(\nabla\times\RR)_\perp}{|\BB|},  
\end{equation}
where the brackets indicate a quantity orthogonal to the magnetic field, i.e. $(\XX)_\perp = \XX - (\bb\cdot\XX)\bb$.  The essential difference between GMR and Eyink's theory arises because GMR assumes laminar ideal MHD outside of an isolated diffusion region (i.e. the magnetic field is frozen into the flow outside $D$) whereas in a turbulent plasma, the frozen-in condition is, in general, violated almost everywhere.

\begin{figure}
    \centering
    \begin{tikzpicture}
        %% Ideal motion
        \draw[thick,->] ([shift=(180:3cm)]-5,0) arc (180:133:3cm);
        \node at (-8,2) {$\rm(a)$};
        \fill[green] (-7.9,0.8)  circle[radius=4pt];
        \fill (-7.9,0.8)  circle[radius=2pt];
        \draw[thick,->] (-7.5,0.75) -- (-5.5,0.75);
        \node at (-6.5,0.7) [anchor=north]{$t\rightarrow t+\delta t$};
        \draw[thick,->] ([shift=(180:3cm)]-2,0) arc (180:133:3cm);
        \fill[green] (-4.9,0.8)  circle[radius=4pt];
        \fill (-5.2,0.8)  circle[radius=2pt];
        %% Slippage
        \draw[thick,->] ([shift=(180:3cm)]1,0) arc (180:133:3cm);
        \node at (-2,2) {$\rm(b)$};
        \fill[green] (-1.9,0.8)  circle[radius=4pt];
        \fill (-1.9,0.8)  circle[radius=2pt];
        
        \draw[thick,->] (-1.5,0.75) -- (0.5,0.75);
        \node at (-0.5,0.7) [anchor=north]{$t\rightarrow t+\delta t$};
        \draw[thick,->] ([shift=(180:3cm)]4.5,0) arc (180:133:3cm);
        \draw[thick,dashed,->] ([shift=(180:3cm)]4.1,0) arc (180:133:3cm);
        \fill[green] (1.2,0.8)  circle[radius=4pt];
        \fill (0.8,0.8)  circle[radius=2pt];
    \end{tikzpicture}
    \caption{A representation of reconnection as field line slippage. A fixed point in space is indicated by a black dot. At time $t$ there is a parcel of plasma (green circle), at this position, that is attached to a field line. (a) shows ideal motion, in which after a small time increment $\delta t$, the parcel remains attached to the field line. Here $\bsigma=\boldsymbol{0}$. (b) shows the same elements but now the parcel of plasma has slipped to a different field line after $\delta t$. Here $\bsigma\ne\boldsymbol{0}$.}
    \label{cartoon}
\end{figure}
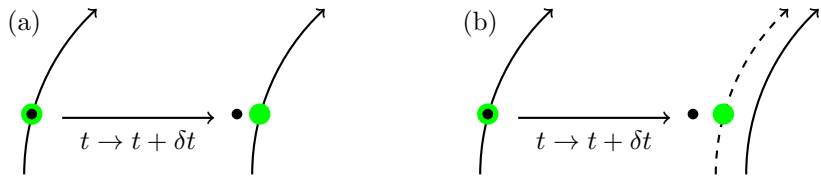
What $\bsigma$ measures is the rate of velocity slip, per unit arclength, along a field line. Although field line velocities are useful tools for understanding fundamental concepts about reconnection, they are not so useful for practical calculations and may not even be well-defined \citep{pontin2022}. There is, however, an alternative and simpler interpretation of equation (\ref{slip_source_velocity}) that does not require field line velocities. To see this, consider a fixed point in space. At a given time, there is a parcel of plasma at this point and a field line passing through it. The relationship between the parcel of plasma and the field line can develop in two distinct ways. First, the parcel of plasma can remain attached to the field line. In other words, the motion is \emph{ideal} and there is no reconnection. Secondly, the parcel could slip relative to the field line. This slippage would require non-ideal terms and satisfies our definition of reconnection (the parcel of plasma is connected to a different field line). This process is illustrated in Figure \ref{cartoon}.

For changes in field line topology, we consider the rate of change of $\bb$ at a fixed point in space. Using the relation $\BB\cdot\BB=|\BB|^2$, it is straightforward to show that
\begin{equation}\label{top_1}
    \frac{\partial\bb}{\partial t} = \frac{1}{|\BB|}\left(\frac{\partial\BB}{\partial t}\right)_\perp.
\end{equation}
It is clear here that we are not choosing the point in space to be a null point, so that we can define a field line passing through this point. Taking the curl of equation (\ref{Ohm_general}), making use of Faraday's law, and substituting this into equation (\ref{top_1}) reveals
\begin{equation}\label{top_2}
    \frac{\partial\bb}{\partial t} =  \frac{1}{|\BB|}(\nabla\times(\uu\times\BB)-\nabla\times\RR)_\perp = \frac{1}{|\BB|}(\nabla\times(\uu\times\BB))_\perp + \bsigma.
\end{equation}
It is, therefore, clear from equation (\ref{top_2}) that $\bsigma$ represents the deviation, from ideal motion, of field line connectivity. In other words, $\bsigma\ne\boldsymbol{0}$ implies the slippage of field lines relative to ideal motion\footnote{This interpretation was also discovered, independently, by \cite{jafari2019}.}, i.e. it represents a change in field line connectivity. Given the above interpretation, a more intuitive name for $\bsigma$ would be the \emph{field line slippage rate}, or just the \emph{slippage rate}. We will use this name throughout this paper.

\subsection{The choice of non-ideal terms}
In order to proceed, we need information about the form of $\RR$, with an eye on applications to the solar corona. The \emph{raison d'\^{e}tre} for the introduction of the slip velocity source in \cite{eyink2015} was to account for reconnection under turbulence. Magnetohydrodynamic (MHD) simulations of active region-scale phenomena are not able to resolve turbulent scales down to the dissipative scale and so require models to represent turbulence. In relation to reconnection, ad hoc models of turbulence are routinely included in such simulations through anomalous resistivity, which typically takes the form of some prescribed function of the current density. An alternative approach, resulting in a more self-consistent model of turbulence, would be to adopt a \emph{renormalization group} (RG) methodology \citep[e.g.][]{McComb1995}. Despite different approaches to RG in MHD turbulence, most begin with some coarse-grained description, through, for example, mollification \citep{eyink2015} or ensemble averaging \citep[][]{Yokoi2020}. For the sake of the discussion, we will focus on the latter.

If we consider an ensemble average of the MHD equations, the individual fields are split into `mean' and `fluctuating' parts. For example, if $f$ is some given field, it is decomposed as $f=\langle f\rangle + f'$, where $\langle f\rangle$ is the mean part (the ensemble average) and $f'$ is the fluctuating part. In the context of turbulence, the ensemble average follows the Reynolds averaging rules \citep{Yokoi2020}. Applying these rules to the ideal induction equation leads to 
\begin{equation}\label{turb_ind}
    \frac{\partial\BB}{\partial t} = \nabla\times(\uu\times\BB + \EE_M),
\end{equation}
where all variables are interpreted as mean quantities and the electromotive force $\EE_M=\langle \uu'\times\BB'\rangle$. Note that equation (\ref{turb_ind}) is equivalent to that formed from mollification, but with a different definition of $\EE_M$ \citep[see equation (2.4) of ][]{eyink2015}. 

One important consequence of ensemble averaging is that when we speak of reconnection in relation to equation (\ref{turb_ind}), it is in relation to the \emph{mean} magnetic field. The electromotive force acts as the source for reconnection rather than any negligible resistivity, which we have ignored here. The corresponding slippage rate, of the mean magnetic field lines, is then
\begin{equation}\label{turb_slip}
    \bsigma = \frac{(\nabla\times\EE_M)_\perp}{|\BB|}.
\end{equation}
It is at this point where a RG approach enters to provide a model for $\EE_M$ in terms of mean variables. One approach that has shown some initial promise as a means to simulating turbulent reconnection in the solar atmosphere, is the two-scale direct-interaction approximation
\citep[TSDIA;][]{yoshizawa2001,Yokoi2020}. This approach has been constructed to model inhomogeneous turbulence, which is necessary for the corona. The application of the TSDIA approach is far too detailed to expand upon here, and the interested reader can consult extensive descriptions in \cite{Yokoi2020} and \cite{yokoi2023}. Instead, for the purposes of this discussion, we will skip straight to the main result, for which the leading terms of the electromotive force can be written as
\begin{equation}\label{emf}
    \EE_M = \alpha_t\BB - \beta_t\jj + \gamma_t\bOmega,
\end{equation}
where $\jj=\nabla\times\BB$ (we set $\mu_0=1$ for convenience); $\bOmega=\nabla\times\uu$ and $\alpha_t$, $\beta_t$ and $\gamma_t$ are turbulent transport coefficients corresponding to the current and kinetic helicities, the turbulent energy and the cross helicity respectively. Each coefficient has its own equation in terms of the mean variables \citep{Yokoi2020}. The terms in equation (\ref{emf}) are also common to other approaches to modelling turbulent MHD, particularly in dynamo theory \citep[][]{brandenburg2005}, with the main differences lying in the specific details of the transport coefficients.

Given that we are primarily concerned with reconnection in the low-$\beta$ corona, it is a reasonable expectation that the dominant term in equation (\ref{emf}) is $-\beta_t\jj$, particularly in regions of strong and dynamically-important reconnection. The term $\alpha_t\BB$ is related to the so-called $\alpha$-effect of dynamo generation \citep[e.g.][]{brandenburg2005}. Thus, this term should play a secondary role in the corona. At  current sheets, where the most dynamically important reconnection takes place, either $\gamma_t$ or $\bOmega$ (or both) are expected to change sign across the current sheet \citep{Yokoi2020}. Thus at the current sheet itself, the term $\gamma_t\bOmega$ is either zero or very small compared to the magnitude of  $-\beta_t\jj$ \citep[see][for some examples of this]{Widmer16a,Widmer19,stanish2024}. Therefore, in the solar corona, a reasonable approximation of the electromotive force is
\begin{equation}\label{approx}
    \EE_M\approx-\beta_t\jj,
\end{equation}
in regions of strong reconnection. This approximation has a very similar mathematical form to the anomalous resistivity used in many MHD simulations of the solar corona, even if the underlying physics is different. We will adopt this form for the non-ideal term for this rest of the paper. However, for ease of reading, we revert to more conventional notation, writing $\eta$ rather than $
\beta_t$ and referring to it as the `resistivity' {(to do this, we equate $-\nabla\times\RR$ with $\nabla\times\EE_M$, together with $\RR=\eta\mathbf{j}$ and equation (\ref{approx}))}. Inserting this choice into equation (\ref{slip_source_velocity}) gives
\begin{equation}\label{slip_gen}
    \bsigma=-\frac{1}{|\BB|}(\nabla\eta\times\jj+\eta\nabla\times\jj)_\perp.
\end{equation}
Equation (\ref{slip_gen}) is straightforward to calculate, given the magnetic field $\BB$, and represents a purely \emph{local} description of reconnection, i.e. it depends only on quantities at the location in space where it is calculated. The results of the rest of this paper are based on equation (\ref{slip_gen}). That being said, the approach does not strictly depend on equation (\ref{slip_gen}) and can easily be adapted to include other terms in the electromotive force. The practical benefit of using equation (\ref{slip_gen}) is that it can be applied directly to current MHD simulations of the solar corona (which do not include $\alpha_t$ or $\gamma_t$) and, as will be explained later, to magnetic field extrapolations from magnetogram observations.

\subsection{Connections to magnetic field geometry}
Although equation (\ref{slip_gen}) is a simple expression, the physical significance of each of its terms is not immediately clear. Progress can be made, however, to relate these terms to the underlying geometry of the magnetic field. To proceed, we follow the approach of \cite{prior2020} by seeking a suitable decomposition of the current density. If we denote the Lorentz force as $\FF=\jj\times\BB$, we can decompose the current density as
\begin{equation}\label{j1}
    \jj = \BB\times\frac{\FF}{|\BB|^2} + \alpha\BB, 
\end{equation}
where 
\begin{equation}
    \alpha = \frac{(\nabla\times\BB)\cdot\BB}{|\BB|^2},
\end{equation}
whose geometric interpretation is the twisting of the magnetic field around the field line passing through the point in space at which we are measuring. The geometric parameter $\alpha$ is related to the local field-aligned current. 

We can also express the first term of equation (\ref{j1}) in a similar way to the second term, i.e.
\begin{equation}\label{j2}
    \jj = \lambda\BB_{f\perp} + \alpha\BB,
\end{equation}
where 
\begin{equation}
    \BB_{f\perp} = \BB\times\frac{\FF}{|\FF|}, \quad \lambda = \frac{(\nabla\times\BB)\cdot\BB_{f\perp}}{|\BB|^2}.
\end{equation}
The geometrical interpretation of $\lambda$ is similar to that of $\alpha$ but now measures the twisting of the magnetic field around $\BB_{f\perp}$, a vector that is orthogonal to both $\BB$ and $\FF$. Further, given $\lambda$, we have that
\begin{equation}
    \FF = \lambda\BB_{f\perp}\times\BB = \lambda|\BB|^2\frac{\FF}{|\FF|}.
\end{equation}
Therefore, $\lambda$ also represents the magnitude of the  Lorentz force relative to the square of the magnetic field strength. For example, if, at a particular point in space, the Lorentz force consisted entirely of the magnetic tension force, we would have $\lambda=|(\bb\cdot\nabla)\bb|$, i.e. a measure of the strength of the curvature of the magnetic field.

Inserting equation (\ref{j2}) into equation (\ref{slip_gen}) gives
\begin{eqnarray}
    \bsigma &=& -\frac{1}{|\BB|}(\CC_1 + \eta\CC_2), \label{sigma_perp}\\
    \CC_1 &=&  \alpha\nabla\eta\times\BB - \lambda(\nabla\eta\cdot\BB)\frac{\FF}{|\FF|}, \nonumber\\ 
    \CC_2 &=& (\lambda\omega_1-\nabla\lambda\cdot\BB)\frac{\FF}{|\FF|} + \lambda(\alpha + \omega_{2})\BB_{f\perp} + \nabla\alpha\times\BB, \nonumber
\end{eqnarray}
where 
\begin{equation}
    \omega_1 = \nabla\times\BB_{f\perp}\cdot\frac{\FF}{|\FF|}, \quad \omega_{2} = \frac{(\nabla\times\BB_{f\perp})\cdot\BB_{f\perp}}{|\BB|^2}.
\end{equation}
In deriving equation (\ref{sigma_perp}), we have decomposed the curl of equation (\ref{j2}) into parts that are parallel and orthogonal to the magnetic field. Only terms orthogonal to $\BB$ appear in the slippage rate, i.e. those parallel to $\FF$ or $\BB_{f\perp}$ or the cross product of $\BB$ with another vector. 

All of the terms in equation (\ref{slip_gen}) relate to specific geometric properties of the magnetic field, and so represent a direct link between field line geometry and slippage. The terms collected into $\CC_1$ depend on a non-uniform resistivity. The first term describes a contribution due to gradients of the resistivity orthogonal to the direction of the magnetic field and weighted by the local field line twist. Here, both the field line geometry and the properties of the plasma (resistivity) combine to produce field line slippage. This term is combined with the second which depends on gradients of resistivity parallel to the magnetic field and acts in the direction of the Lorentz force.

The terms collected into $\CC_2$ depend only on the field line geometry. The first term of $\CC_2$, in the direction of the Lorentz force, combines the rotation of the magnetic field around the Lorentz force with gradients of the relative strength of the Lorentz force along field lines. The second term, orthogonal to both the magnetic field and the Lorentz force, combines local winding and Lorentz forces together with the rotation of these vectors in a direction orthogonal to both (that of $\BB_{f\perp}$). The third term describes the contribution of gradients in the field-aligned winding orthogonal to the direction of the magnetic field.

\subsection{The force-free limit}

In the low plasma-$\beta$ environment of the solar corona, the force-free approximation, $\nabla\times\BB=\alpha\BB$, is often adopted. In terms of the analysis we have presented, this coincides with $\FF=\boldsymbol{0}$. Taking this approximation simplifies the analysis significantly, resulting in
\begin{equation}
    \bsigma = -\nabla(\alpha\eta)\times\bb.
\end{equation}
It is useful to explore some particular examples.

For a linear force-free (LFF) field ($\alpha = $ const.), we have
\begin{equation}
    \bsigma_{\rm LFF} = -\alpha\nabla\eta\times\bb.
\end{equation}
First, if $\eta$ is a constant, then there is no field line slippage for LFF fields, which is a well-known result \citep[][]{JETTE1970,seehafer1993}. If $\alpha\ne0$ (non-potential), however, then there is the possibility of field line slippage, though this depends on the specific form of $\eta$.

For a constant resistivity, a nonlinear force-free (NLFF) field is required as field line slippage can only occur if there are gradients in the field-aligned currents. We denote this slippage rate as
\begin{equation}\label{slip_nlff}
    \bsigma_{\rm NLFF} = -\eta\nabla\alpha\times\bb.
\end{equation}
Equation (\ref{slip_nlff}) is particularly useful as it can be applied to NLFF extrapolations. We will discuss a particular example shortly.

\section{Illustrative examples}
We now present some applications of the slippage rate analysis to coronal magnetic fields. 

\subsection{Analytical flux rope formation}
\begin{figure}    %%%%%%%%%%%%%%%%%% FIGURE 2
                          % includes the two top panels 
\centerline{\hspace*{0.015\textwidth}
         \includegraphics[width=0.515\textwidth,clip=]{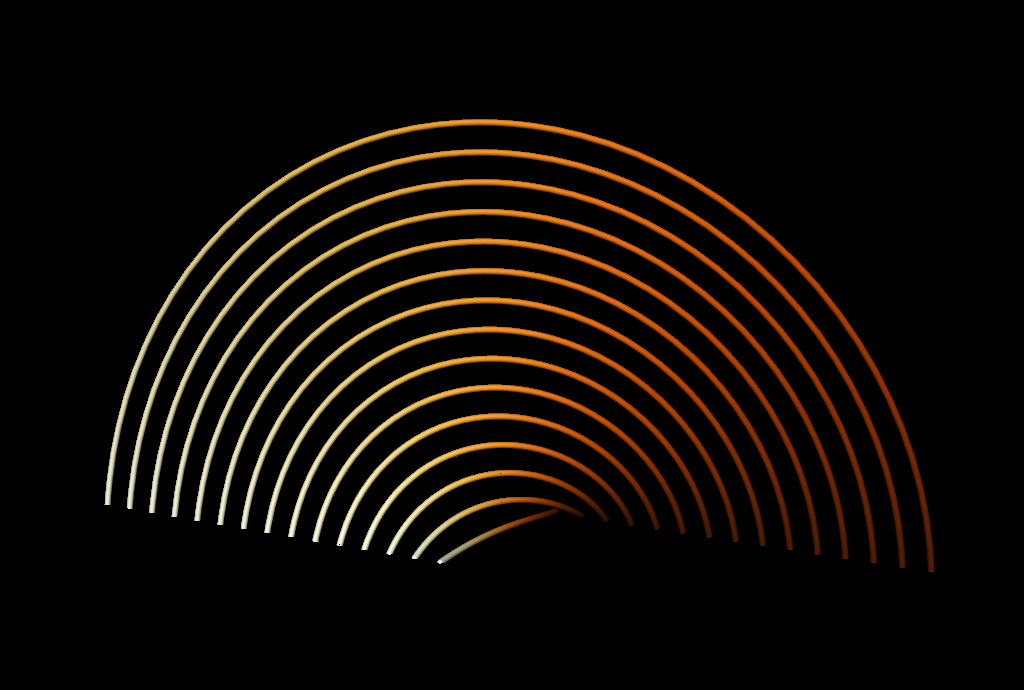}
         \hspace*{-0.03\textwidth}
         \includegraphics[width=0.515\textwidth,clip=]{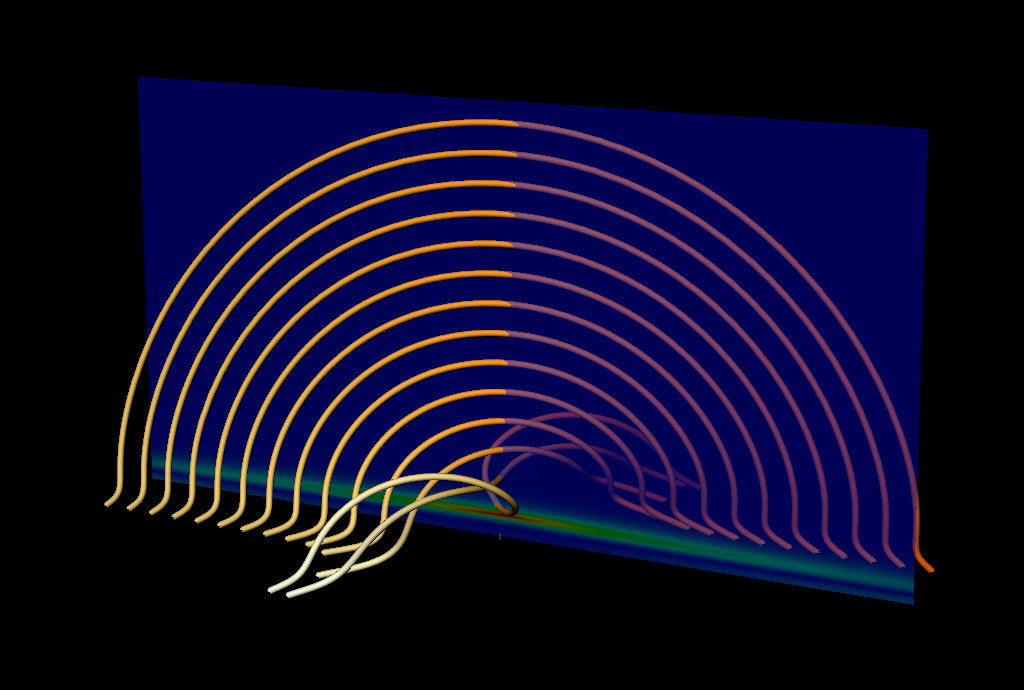}
        }
\vspace{-0.33\textwidth}   % Shift close to the panel top 
\centerline{\large \bf     % Includes the labels (here needs the color 
                          %   package, see beginning of this file)
\hspace{0.0 \textwidth}  \color{white}{(a)}
\hspace{0.415\textwidth}  \color{white}{(b)}
   \hfill}
\vspace{0.31\textwidth}    % Shift back to the panel bottom 
\centerline{\hspace*{0.015\textwidth}
         \includegraphics[width=0.515\textwidth,clip=]{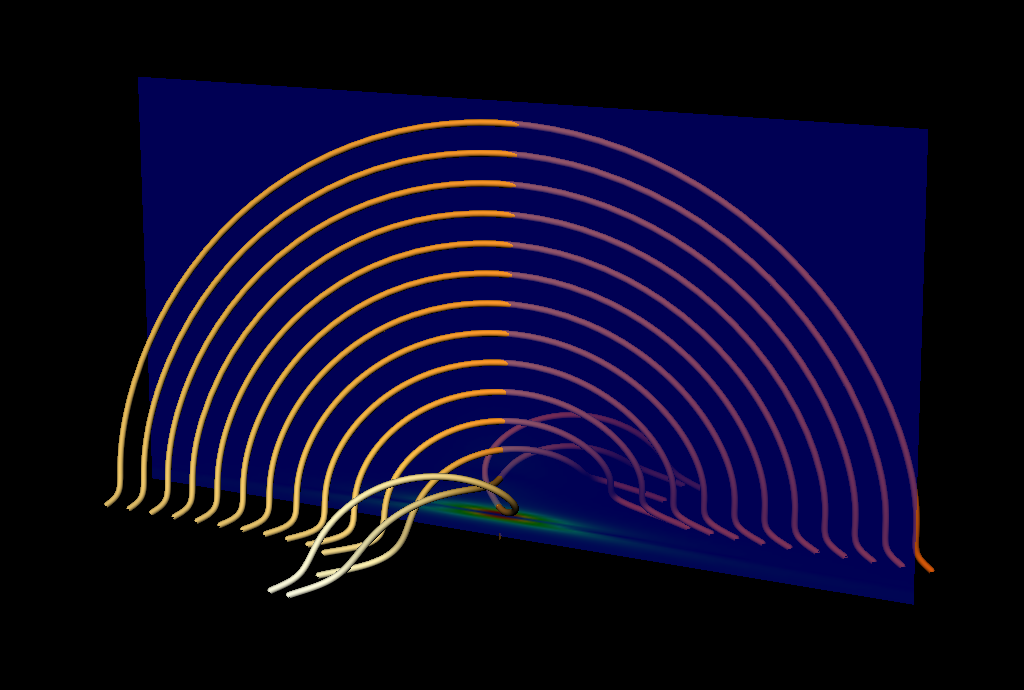}
         \hspace*{-0.03\textwidth}
         \includegraphics[width=0.515\textwidth,clip=]{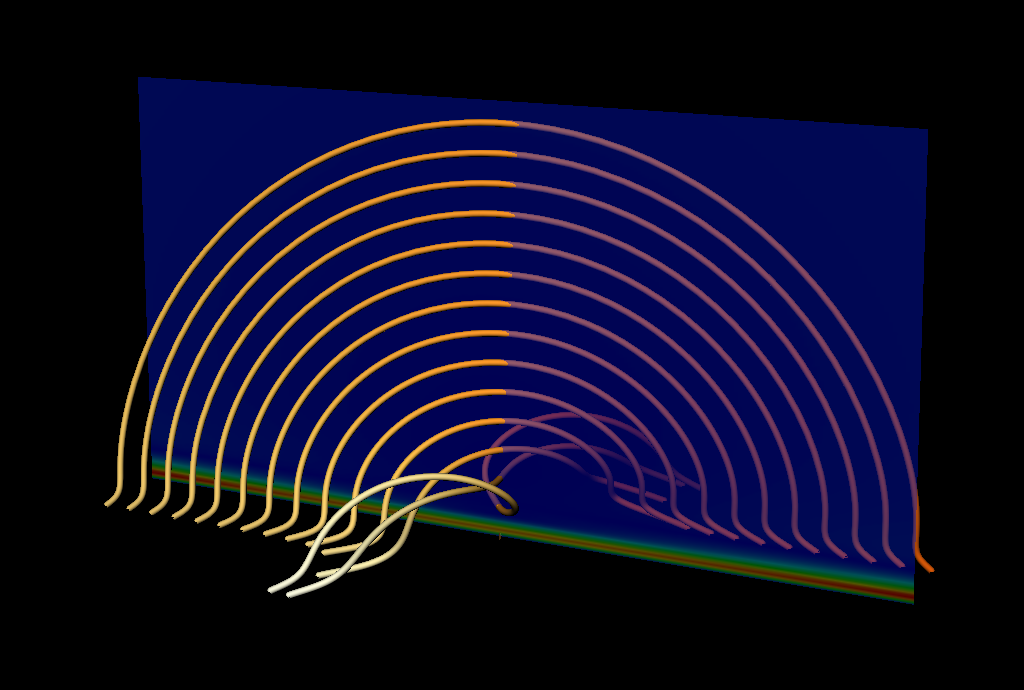}
        }
\vspace{-0.33\textwidth}   % Shift close to the panel top 
\centerline{\large \bf     % Includes the labels (here needs the color package)
\hspace{0.0 \textwidth} \color{white}{(c)}
\hspace{0.415\textwidth}  \color{white}{(d)}
   \hfill}
\vspace{0.31\textwidth}    % Shift back to the panel bottom 
              
\caption{Visualizations of magnetic field lines at (a) $t=0$ and (b-d) $t=40$. The field lines are identical in (b-d) but the variable on the vertical slice changes. (b) displays $|\bsigma|$, (c) $|\bsigma_{\rm NLFF}|$ and (d) $|\FF|$. The colour map from red to blue indicates strong to weak.
        }
\label{toy}
\end{figure}

As a first demonstration, we consider the kinematic model studied in \cite{hesse2005} and \cite{titov2009}. This model is a simple analytical model of flux rope formation and has been studied in the above works through GMR and QSL analyses respectively. It is based on a magnetic field with the components\footnote{These components are modified compared to the original ones presented in \cite{hesse2005}. This representation was presented by Anthony Yeates and Gunnar Hornig at the Royal Astronomical Society National Astronomy Meeting 2013.}

\begin{eqnarray*}
    B_x &=& 3, \\
    B_y &=& -\left(z+\frac{(1-z^2)t}{(1+z^2)^2(1+x^2/36)}\right), \\
    B_z &=& y,
\end{eqnarray*}
where $t$ is a parameter representing time. At $t=0$, there is a sheared arcade with a constant current density $\jj = 2\ee_x$. Therefore, the magnetic field is not force-free and so it may be expected that terms involving both $\lambda$ (related to the Lorentz force) and $\alpha$ (related to field-aligned currents) may provide important contributions to field line reconnection (we assume here that $\eta$ = const. for convenience). A visualization of the initial arcade is shown in Figure \ref{toy} (a).

Given that the current density is constant at $t=0$, there is no immediate reconnection. As $t$ increases, the field lines evolve in a non-ideal way. That is, in order to move from one time to another, the magnetic field must undergo reconnection. As soon as $t>0$, we have $\bsigma\ne\boldsymbol{0}$. However, in order to reveal some of the reconnection properties in more detail, let us skip to $t=40$, once a flux rope has developed. Field lines at this time are traced in Figure \ref{toy} (b-d), together with vertical slices displaying different variables that we now discuss.

In Figure \ref{toy} (b), the central slice displays $|\bsigma|$ (red is strong and blue is weak). What is shown is that the field line slippage rate dominates in the centre, just under the flux rope, but is also having an effect in a horizontal band structure that stretches across the slice. Before discussing this further, let us consider the behaviour of $\bsigma_{\rm NLFF}$ which is shown in Figure \ref{toy} (c). Now since $\BB$ is not a NLFF field, this variable is only one of the components making up the slippage rate (here we make use of the symbol `$\bsigma_{\rm NLFF}$' so as not to introduce too much unnecessary notation). It reveals the contribution to slippage of gradients of local twist. As can be seen from Figure \ref{toy} (c), $|\bsigma_{\rm NLFF}|$ is concentrated in the centre where $\max|\bsigma|$ occurs. It also exhibits more structure compared to $|\bsigma|$, but does not stretch across the full horizontal extent. This can only mean that the terms in equation (\ref{sigma_perp}) related to the Lorentz force must play an important role in contributing to slippage rate across the entire band. In Figure \ref{toy} (d), the slice shows the magnitude of the Lorentz force. This stretches in a band across the entire slice in a similar way to $|\bsigma|$. This behaviour is to be expected given the curvature of the field lines in this region that indicate magnetic tension. In this analysis, both the presence of Lorentz forces and gradients in field-aligned currents play a role in the behaviour of reconnection at different locations. Whilst the strongest reconnection occurs under the flux rope, field line slippage also occurs across the lower part of the sheared arcade where the field lines are under tension, i.e. where their curvature has an opposite sign to that of the large-scale arcade.

\begin{figure}    %%%%%%%%%%%%%%%%%% FIGURE 1 
\centerline{\includegraphics[width=0.8\textwidth,clip=]{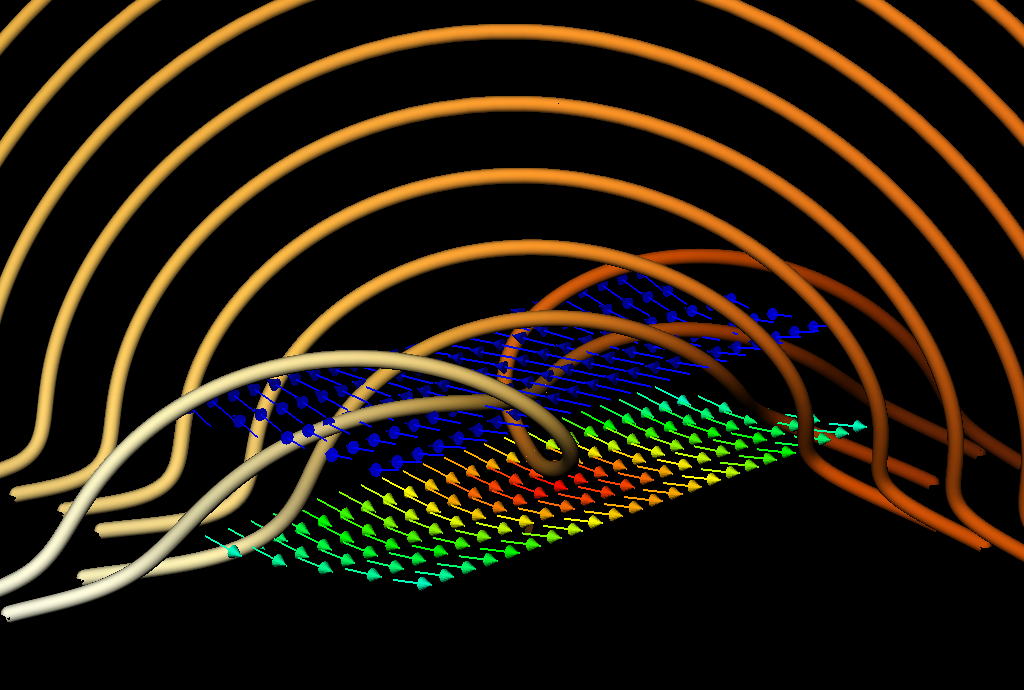}}
\small
        \caption{A visualization of $\bsigma$ (arrows) at the flux rope at $t=40$. The directions of $\bsigma$ are displayed just below and within the flux rope. Magnitude is indicated by colour, red: strong; blue: weak.}
\label{vector}
\end{figure}

As well as the magnitude of the slippage rate, we may also determine its direction. For this example, the vector of the slippage rate rotates with height. A demonstration of this is shown in Figure \ref{vector}, in which the slippage rate vector is shown in two surfaces above and below the flux rope at $t=40$ (the colour scheme is the same as that of the slice in Figure \ref{toy} (b)).

The flux rope in this toy model is created by the time-dependent term in $\BB_y$. One effect of this perturbation is the shearing of field lines. If the time evolution of $\BB$ were an ideal motion, the field lines would continue to shear, with the consequential build-up of current density at the centre of the shear region. Instead, reconnection occurs and leads to the formation of a flux rope. The flux rope field lines are, in places, close to being orthogonal to the direction of the overlying sheared arcade. What the slippage rate vector is indicating here is how the field line motion is deviating from ideal motion, i.e. field line connectivity is slipping in a direction orthogonal to the direction of the overlying sheared arcade field lines in order to form the flux rope field lines. This process is strongest in a particular location that corresponds to the base of the flux rope, i.e. the reconnection region where the flux rope is formed.

\subsection{NLFF extrapolation of AR11158}
We now consider an example based on solar data - the NLFF extrapolation of a vector magnetogram. It was shown earlier that the expression for the slippage rate assuming a NLFF field, equation (\ref{slip_nlff}), is greatly simplified compared to when a non-zero Lorentz force is present. Before calculating $\bsigma_{\rm NLFF}$, however, it is worth considering what this quantity means in this context, since a NLFF extrapolation is a static equilibrium. When a NLFF extrapolation of an active region magnetogram (or larger patch of the solar surface) is performed, it is assumed that the resulting magnetic field represents a good approximation of both the large-scale geometry and topology of the magnetic field. The fact that it is an equilibrium is a consequence of its construction. However, the question may be asked, how would this equilibrium change if there was an instantaneous change from the condition $\partial\BB/\partial t = \boldsymbol{0}$ to $\partial\BB/\partial t = -\nabla\times\RR$? In this way, we may make use of the slippage rate to show where reconnection would occur and where it would be strongest.

A typical way to analyze the magnetic topology of NLFF extrapolations is to apply QSL analysis \citep[e.g.][]{masson2017,KAIFENG2018,zhao2014,jing2024}. In order to make a comparison with such an analysis, we consider the NLFF extrapolation of AR11158, whose magnetic topology has been studied, using QSL analysis, in \cite{zhao2014}. Our purpose here is not to study the evolution of AR11158, which has been treated in many other works \citep[e.g.][]{schrijver2011,sun2012,chinztzolou2019}. Rather, our aim here is to present an example of what information can be gathered from the calculation of $\bsigma_{\rm NLFF}$ and how this compares qualitatively to what has been found through QSL analysis. Again, for simplicity, we will assume that $\eta$ is constant for this demonstration.

Here we focus on the magnetic structure of the active region shortly before the onset of a CME associated with an X2.2 flare, occurring at approximately 01:24 UT on 15/02/2011 \citep[][]{aslam2024}. In the topological analysis of \cite{zhao2014}, they investigate the formation of a \emph{hyperbolic flux tube} (HFT) which is defined by the intersection of QSLs and has been studied in several works. In the left panel of Figure 9 of \cite{zhao2014}, they show a map of the squashing factor \citep{pariat2012} in a cross-sectional cut across the HFT, at 23:58 UT on 14/02. The characteristic `teardrop' shape of a HFT is clearly visible, together with some complex internal structure. The main inference from this QSL configuration is that there is a strongly twisted flux rope, which later erupts as a CME. We will now consider what can be discovered by analyzing the same event using $\bsigma_{\rm NLFF}$.

In order to create a NLFF extrapolation of AR11158, we make use of the deep learning method of \cite{jarolim2023}. Since the extrapolation is based on  Space-Weather Helioseismic and Magnetic Imager (HMI) Active Region Patches (SHARP) vector magnetograms \citep[][]{2hoeksma014}, the nearest time available to us, in relation to that studied by \cite{zhao2014}, is at 00:00 UT on 15/02, which is just a difference of two minutes. Once the magnetic field is produced, we calculate $\bsigma_{\rm NLFF}$ with second-order finite differences. 

\begin{figure}    %%%%%%%%%%%%%%%%%% FIGURE 3
                          % includes the two top panels 
\centerline{\hspace*{0.015\textwidth}
         \includegraphics[width=0.515\textwidth,clip=]{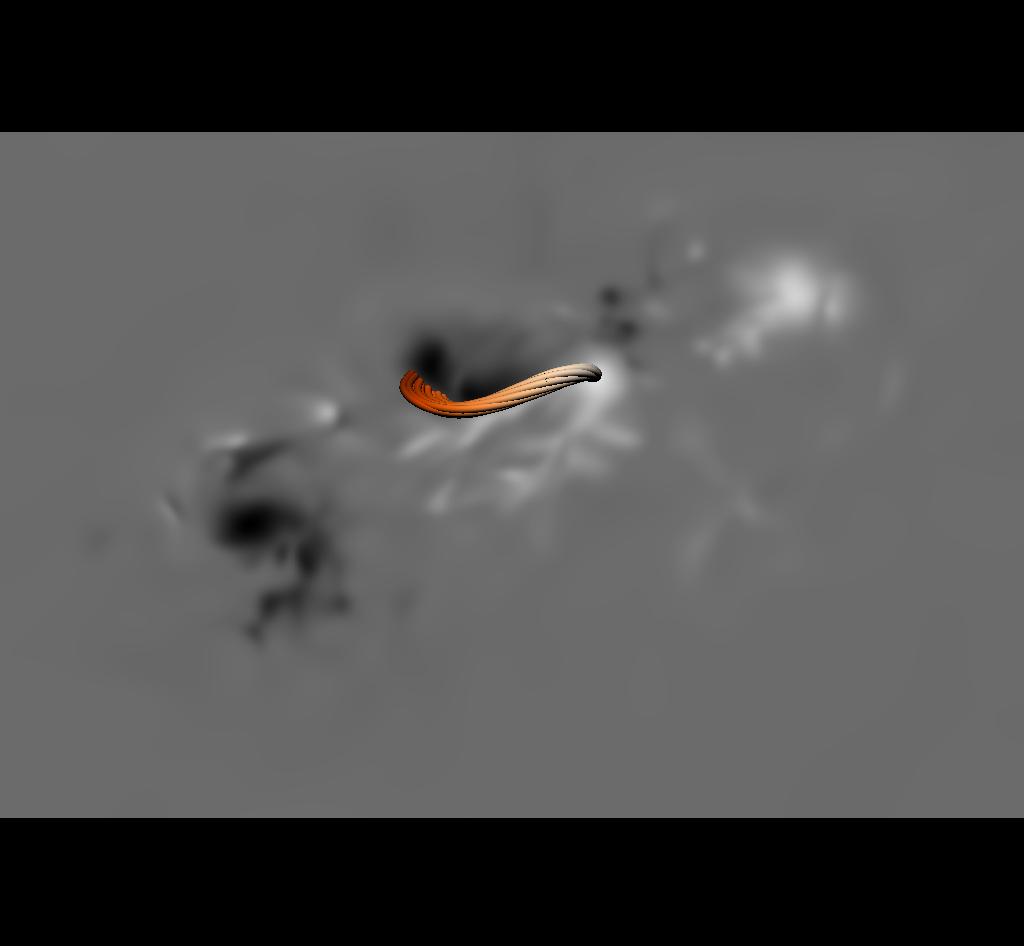}
         \hspace*{-0.03\textwidth}
         \includegraphics[width=0.515\textwidth,clip=]{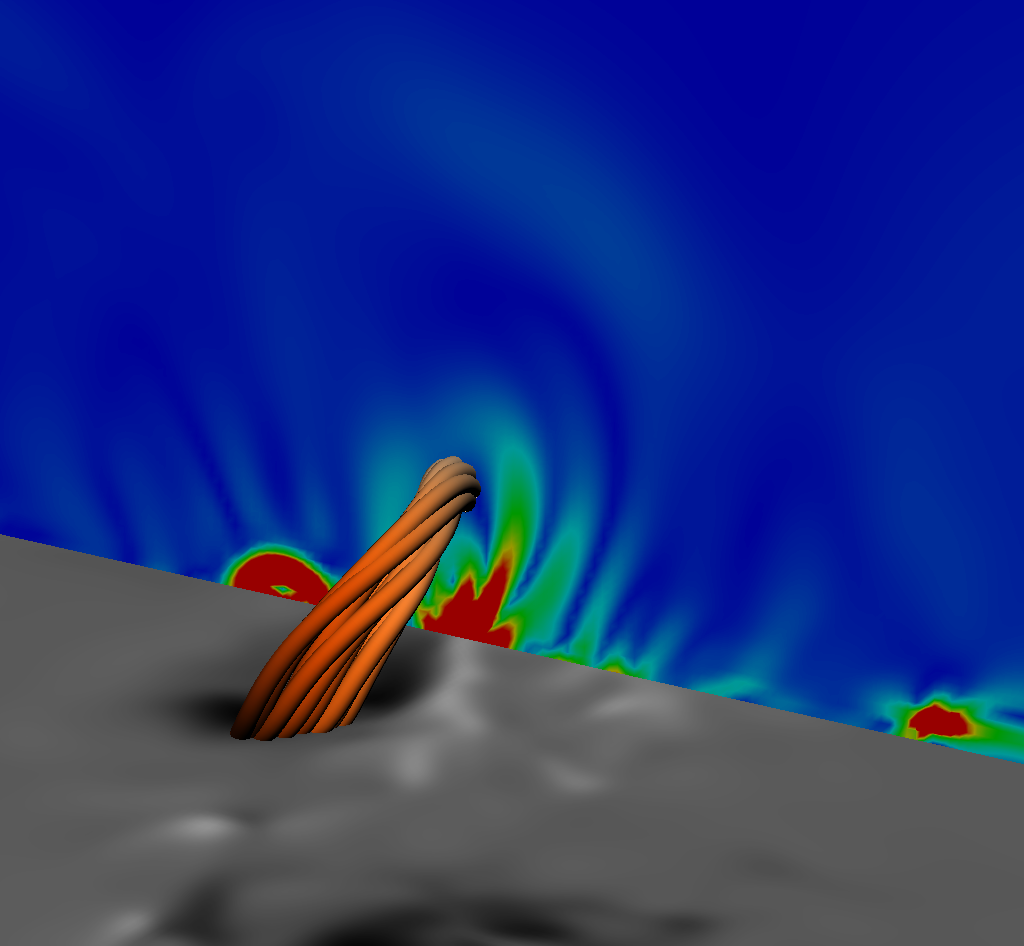}
        }
\vspace{-0.46\textwidth}   % Shift close to the panel top 
\centerline{\large \bf     % Includes the labels (here needs the color 
                          %   package, see beginning of this file)
\hspace{0.0 \textwidth}  \color{white}{(a)}
\hspace{0.415\textwidth}  \color{white}{(b)}
   \hfill}
\vspace{0.41\textwidth}    % Shift back to the panel bottom 
\
%\vspace{0.31\textwidth}    % Shift back to the panel bottom 
              
\caption{A flux rope traced using the strucutre of the slippage rate magnitude. (a) shows the twisted flux rope along the central PIL of the active region (the {magnetogram} shows $B_z$, white: positive, black: negative). (b) show the flux rope and the slice of $|\bsigma_{\rm NLFF}|$ from which the rope is traced. Red to blue indicates strong to weak $|\bsigma_{\rm NLFF}|$.   
        }
\label{nlff_mag_and_S}
\end{figure}

Figure \ref{nlff_mag_and_S} (a) displays a magnetogram of $B_z$ of AR11158 at 00:00 UT on 15/02. At the central polarity inversion line (PIL), field lines are traced and reveal a twisted flux rope. The field lines of this rope have been traced from a location determined by the behaviour of $\bsigma_{\rm NLFF}$, which is shown in a vertical slice in Figure \ref{nlff_mag_and_S} (b). The slice lies along a north-south direction cutting the central PIL. There are several arcade-like structures of $|\bsigma_{\rm NLFF}|$ with stronger slippage rates near the photosphere. One of these structures is closed, forming a ring shape of non-zero slippage rate (strong at the base and weaker at the top) with little or no slippage in the centre. The field lines of the flux are traced from the central region of no slippage. This structure is similar in shape to the teardrop structure of the HFT found by \cite{zhao2014}.

As well as detecting a HFT-like structure, more information can be found from the $|\bsigma_{\rm NLFF}|$ map. The reconnection signature is asymmetric, with a stronger slippage rate on one side. We now investigate the relationship of this stronger patch of slippage to the magnetic field. 

\begin{figure}    %%%%%%%%%%%%%%%%%% FIGURE 2
                          % includes the two top panels 
\centerline{\hspace*{0.015\textwidth}
         \includegraphics[width=0.515\textwidth,clip=]{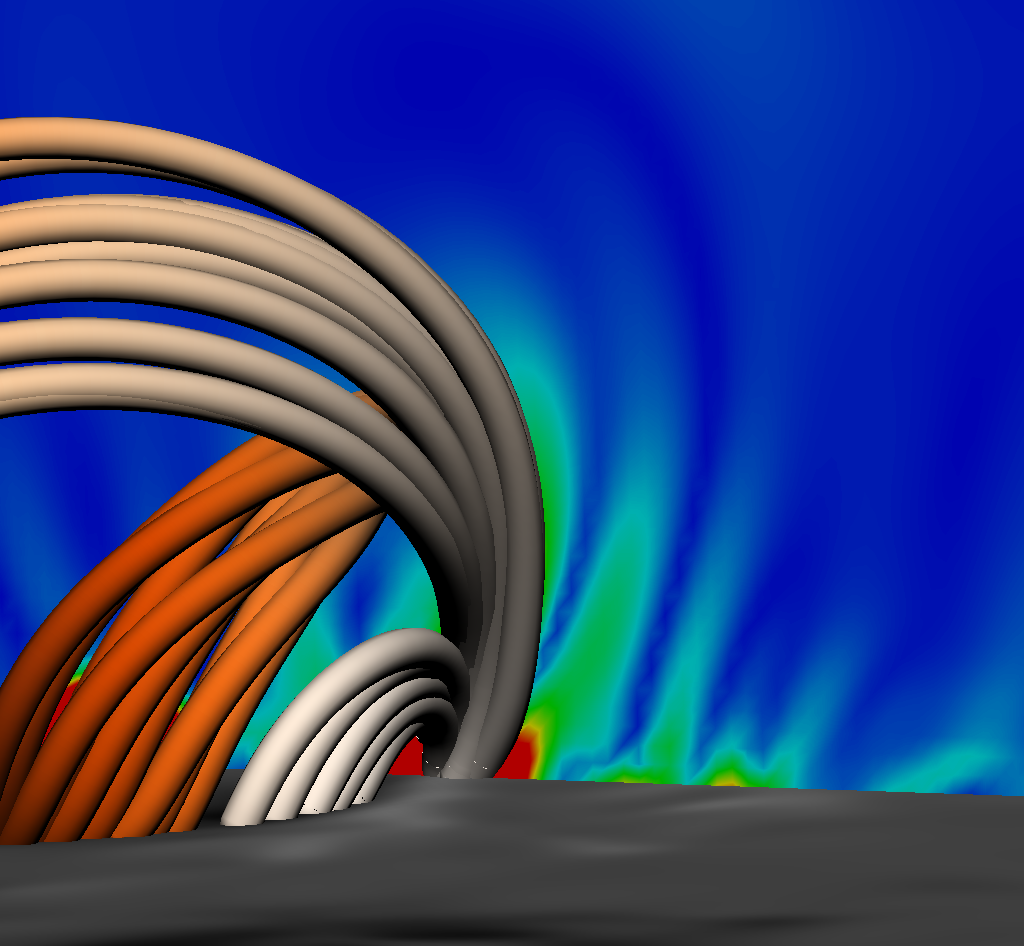}
         \hspace*{-0.03\textwidth}
         \includegraphics[width=0.515\textwidth,clip=]{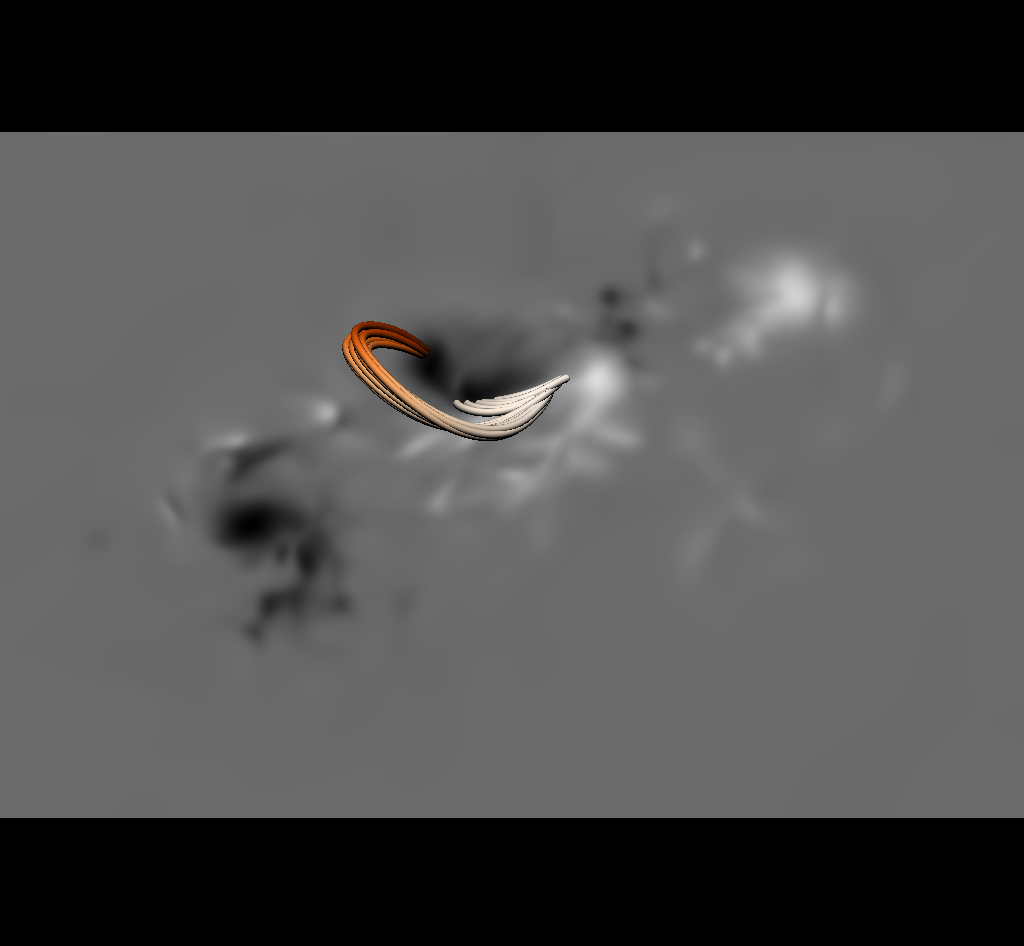}
        }
\vspace{-0.46\textwidth}   % Shift close to the panel top 
\centerline{\large \bf     % Includes the labels (here needs the color 
                          %   package, see beginning of this file)
\hspace{0.0 \textwidth}  \color{white}{(a)}
\hspace{0.415\textwidth}  \color{white}{(b)}
   \hfill}
\vspace{0.41\textwidth}    % Shift back to the panel bottom 
\centerline{\hspace*{0.015\textwidth}
         \includegraphics[width=0.515\textwidth,clip=]{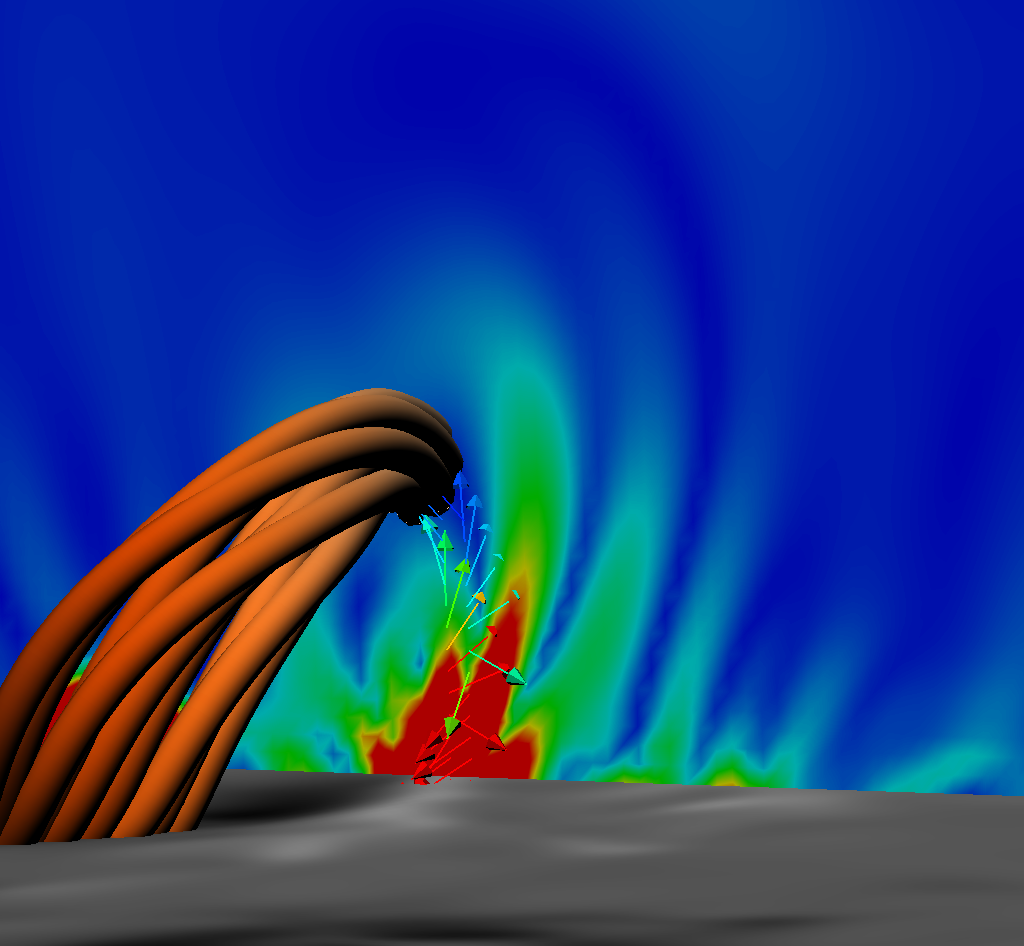}
         \hspace*{-0.03\textwidth}
         \includegraphics[width=0.515\textwidth,clip=]{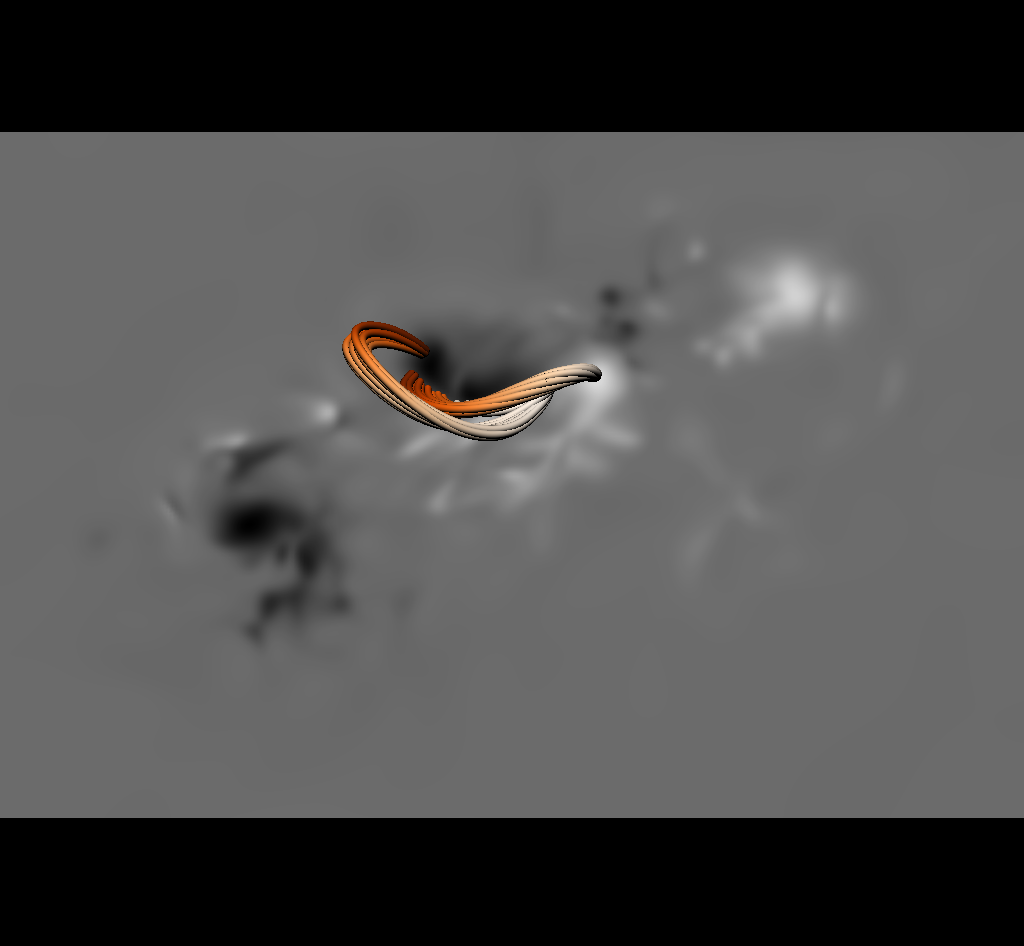}
        }
\vspace{-0.46\textwidth}   % Shift close to the panel top 
\centerline{\large \bf     % Includes the labels (here needs the color package)
\hspace{0.0 \textwidth} \color{white}{(c)}
\hspace{0.415\textwidth}  \color{white}{(d)}
   \hfill}
\vspace{0.41\textwidth}    % Shift back to the panel bottom 
              
\caption{Relating the stronger slippage rates to field line structure. (a) shows two `loops' of field lines traced from the patch of strong slippage on the slice, together with the flux rope. (b) shows these loops on the magnetogram without the flux rope. (c) shows the same as (a) but with the loops removed and replaced by a visualization of the vector $\bsigma_{\rm NLFF}$. (d) shows the two loops and the flux rope on the magnetogram. Colour conventions are the same as in Figure \ref{nlff_mag_and_S}.
        }
\label{nlff_vec}
\end{figure}

Figure \ref{nlff_vec} (a) displays a similar image to that of Figure \ref{nlff_mag_and_S} (b), but now with extra field lines traced from the region of strong slippage rate shown by the red patch on the vertical slice. The magnetic field has the form of two magnetic loops - one short and low-lying and the other much larger. Figure \ref{nlff_vec} (b) shows the horizontal extent of these loops, with the smaller loop straddling the PIL and connecting to the central positive patch of $B_z$, and the larger loop connecting down at the PIL but also the western edge of the central negative patch of $B_z$. In order to better understand how the slippage rate relates to this magnetic structure, we can look at the direction of $\bsigma$, which is shown in Figure \ref{nlff_vec} (c). Nearest to the photosphere, the slippage rate is directed along the direction of the PIL. This is where the two loops meet on the PIL.  Moving higher in the atmosphere, the direction of the slippage rate rotates and becomes directed toward the flux rope. This is indicative of reconnection between the two loops, sheared along the PIL, reconnecting to contribute to the flux rope. We do not claim that is is how the flux rope formed - answering this would require a detailed study of the evolution of the magnetic field in time, a task which goes beyond the scope of this paper. However, what has been shown is that reconnection at this time would occur between the two identified loops causing a field line slippage rate to be directed toward the flux rope. Further, the slippage rate at the flux rope itself is very weak, suggesting that this is a stable structure (at least at this time). Figure \ref{nlff_mag_and_S} (d) shows the information of (b) including the flux rope. These results agree qualitatively with the identification of a flux rope using QSL analysis, by \cite{zhao2014}.

\section{Summary and discussion}
The purpose of this work has been to introduce a {new description} of reconnection based entirely on local information about the plasma. That is, in contrast to the theory of General Magnetic Reconnection (GMR) and Quasi-Separatrix Layer (QSL) analysis, this {approach} does not require information integrated along field lines across the magnetic field under study. {This new description} combines a re-interpretation of the slip velocity source of \cite{eyink2015} and the geometric analysis of \cite{prior2020}. The result is a {description} that provides information on the location, strength and direction of the field line slippage rate, together with the key geometric features of the magnetic field which contribute to field line slippage. In particular, the geometry of the magnetic field is characterized by the Lorentz force and field-aligned currents, thus linking the geometry to fundamental physical properties of magnetic fields in the solar corona.

We have argued on physical grounds that, for applications in the solar corona, the anomalous resistivity employed in many MHD simulations represents a reasonable approximation of the dominant term in the electromotive force, derived from an averaged form of the MHD equations to model turbulence. In our discussion, we focus on a particular methodology, the TSDIA approach. However, the result is not strongly dependent on the particular turbulence model, all of which produce similar expansions for the leading terms of the electromotive force. 

If a more detailed simulation were to be performed, which included additional terms in the electromotive force or, indeed, extra terms in a generalized Ohm's law, the main elements of connecting the slippage rate to the underlying field line geometry are the same as we have described here. For example, let us focus on the contribution of the $\alpha$-effect term to the slippage rate. A simple calculation yields
\begin{equation}
    \bsigma = \frac{1}{|\BB|}[\nabla\alpha_t\times\BB + \alpha_t\lambda\BB_{f\perp} + \dots].
\end{equation}
Thus, if the $\alpha$-effect term were to be included, the slippage rate would depend on gradients in the transport coefficient $\alpha_t$ (first term) and the relative strength of the Lorentz force (second term).

We have illustrated {our new approach} in two applications related to flux ropes in the solar corona. The first is a modified form of the analytical model introduced by \cite{hesse2005}. We have shown how the deformation of a sheared arcade leads to field line slippage in relation to the magnetic field geometry. The second is an application of the {approach} to a nonlinear force-free extrapolation. Ignoring the Lorentz force significantly simplifies the expression of the field line slippage rate which, assuming constant resistivity, acts orthogonal to the magnetic field direction and depends on gradients of field-aligned twist. We compare with the QSL analysis of \cite{zhao2014}, examining the behaviour of the magnetic field of AR11158 before the onset of a CME. Applied to a static and ideal equilibrium, the results are to be interpreted as what field line slippage would happen if the given equilibrium were the instantaneous state of the magnetic field governed by an Ohm's law with non-ideal terms. We show that the results of our analysis and those of \cite{zhao2014} match qualitatively in that they both indicate the presence of a well-defined pre-CME twisted flux rope. 

{Our description of reconnection,} connecting the local field line slippage rate to the underlying geometry of magnetic field lines, represents a powerful tool for the analysis of active regions dynamics in the solar corona, both in simulations and extrapolations, and is adaptable to both laminar and turbulent reconnection.

\begin{acks}[Acknowledgements]
The author would like to acknowledge support from a Leverhulme Trust grant (RPG-2023-182), a  Science and Technologies Facilities Council (STFC) grant (ST/Y001672/1) and a Personal Fellowship from the Royal Society of Edinburgh (ID: 4282). 
\end{acks}

     % format of references provided by the journal (.bst)
\bibliographystyle{spr-mp-sola}
     % name your Bibtex file containing your references (.bib)
\bibliography{slip}

\end{document}